\documentclass[sip,biber]{now-journal} % Replace example with the journal code.

\usepackage[utf8]{inputenc}
\usepackage{amssymb}
\usepackage{float}

\usepackage{arydshln}

\usepackage{wrapfig}

\title{Multi-Step Prediction and Control of Hierarchical Emotion Distribution in Text-to-Speech Synthesis}

\author[1,3]{Sho Inoue}
\author[4]{Kun Zhou}
\author[2]{Shuai Wang}
\author[1,3,5*]{Haizhou Li}

\affil[1]{School of Data Science, The Chinese University of Hong Kong, Shenzhen (CUHK-Shenzhen), China}

\affil[2]{School of Intelligence Science and Technology, Nanjing University, Suzhou, China}

\affil[3]{Shenzhen Research Institute of Big Data, Shenzhen, China}

\affil[4]{Tongyi Speech Lab, Alibaba Group, Singapore}
\affil[5]{Department of ECE, National University of Singapore, Singapore}

\issuevolumeyear{2024}
\issuevolumenumber{xx}
\issueelocation{e} % This is only required for sip articles
\articledatabox{Received xx xxxxx 2024; Revised xx xxxxx 2024\\[2pt]
ISSN 2048-7703; DOI 10.1561/116.xxxxxxxx\\
\copyright\ 2025 S. Inoue}

\OAstatement{This is an Open Access article, distributed under the terms of the Creative Commons Attribution licence (\url{http://creativecommons.org/licenses/by-nc/4.0/}), which permits unrestricted re-use, distribution, and reproduction in any medium, for non-commercial use, provided the original work is properly cited.}


\creditline*{Corresponding author: haizhouli@cuhk.edu.cn}

\begin{document}

\begin{abstract}
We investigate hierarchical emotion distribution (ED) for achieving multi-level quantitative control of emotion rendering in text-to-speech synthesis (TTS). We introduce a novel multi-step hierarchical ED prediction module that quantifies emotion variance at the utterance, word, and phoneme levels. By predicting emotion variance in a multi-step manner, we leverage global emotional context to refine local emotional variations, thereby capturing the intrinsic hierarchical structure of speech emotion.  Our approach is validated through its integration into a variance adaptor and an external module design compatible with various TTS systems.  Both objective and subjective evaluations demonstrate that the proposed framework significantly enhances emotional expressiveness and enables precise control of emotion rendering across multiple speech granularities.
%We propose a sequential prediction framework that predicts hierarchical ED from textual cues at the utterance, word, and phoneme levels, thereby capturing the inherent hierarchical structure of speech emotion. We investigate two strategies to integrate ED into text-to-speech (TTS) systems: a variance adaptor within the FastSpeech2 framework and an external module compatible with any TTS model. Our design also provides an intuitive, quantitative method to manipulate emotion at multiple granularities, allowing users to adjust emotional intensity during synthesis. Objective and subjective evaluations confirm that our framework enhances text-based emotion rendering and overall speech performance.

\end{abstract}

%\begin{figure*}
%\centerline{\includegraphics[width=1.0\columnwidth]{images/inoue1.png}}
%\caption{(a) Model architecture with hierarchical emotion distribution (hierarchical ED) mechanism in TTS. During inference, the framework extracts hierarchical ED from input audio (``emotion editing''). Users can manually modify hierarchical ED to control emotion intensities at phoneme, word, and utterance levels; 
%(b) Hierarchical ED extraction workflow for emotion distributions at phoneme, word, and utterance levels. 
%}
%\label{fig:training}
%\vskip -1em
%\end{figure*}

\section{Introduction}
Text-to-speech (TTS) synthesis focuses on generating human-like speech from text input \cite{tan2021survey}. Advancements in deep learning have significantly improved the naturalness and quality of synthesized speech. However, current TTS systems still struggle with conveying emotional expressiveness and precisely controlling emotional nuances, limiting their ability to deliver humanlike expressive speech \cite{triantafyllopoulos2023overview}. To address these limitations, Emotional TTS aims to bridge this gap by enhancing speech expressiveness, enabling more engaging and empathetic dialogue systems with emotional intelligence \cite{triantafyllopoulos2024expressivity}.

Emotional TTS faces challenges stemming from the hierarchical structure of human emotions \cite{kun2022emotion}.
Since speech emotion is characterized by distinct prosodic patterns at the phoneme, word, and utterance levels~\cite{triantafyllopoulos2023overview, kun2022emotion, hirschberg2006pragmatics}, these patterns naturally form a hierarchy, as established in previous studies~\cite{el2011survey, schuller2018speech}. The prior literature indicates that modifying only global prosodic attributes does not capture the full complexity of emotional speech~\cite{zhou2023mixedevc, xu2011speech, latorre2008multilevel, zhou2020transforming}. Additionally, prior work in text-to-speech synthesis and emotional voice conversion underscores the necessity of multi-level modeling~\cite{ming2016deep, zhou2020transforming, lei2022multiscale}. Consequently,  developing a method to model the hierarchical structure of emotions is essential for generating nuanced speech synthesis. However, existing text-based emotion representation prediction networks in controllable models address phoneme-level variations~\cite{fastspeech2, msemotts}, overlooking the benefits of multi-level emotion modeling.

\begin{figure}[h!]
\centerline{\includegraphics[width=1.0\columnwidth]{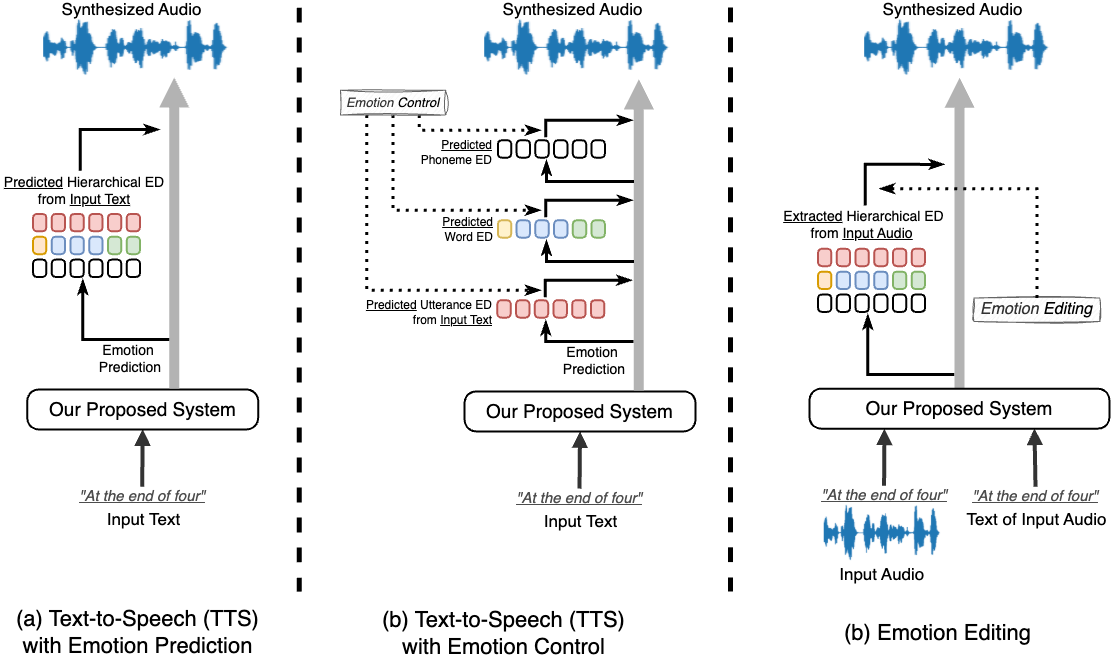}}
\caption{
Inference diagram of the proposed system: (a) TTS with emotion prediction; (b) 
 TTS with emotion control; and (c) emotion editing. The hierarchical emotion distribution (ED) can be obtained in three ways: (1) directly predicted from the input text (``Emotion Prediction''), (2) predicted from the input text with user modifications (``Emotion Control''), or (3) extracted from the input audio and manually adjusted by users (``Emotion Editing'').%In the Emotion Editing mode, we expect inputs of audio and corresponding transcriptions, and users manually edit the hierarchical emotion distribution (ED). In the Emotion Control mode, we require only text input; our system then sequentially predicts the utterance-, word-, and phoneme-level EDs. Users may modify the EDs at any step.
}
\label{fig:intro}
\end{figure}

In this work, we build upon our previous work on a multi-level quantifiable method for speech emotion control~\cite{ShoICASSP} and editing~\cite{ShoAPSIPA} by proposing a multi-step prediction framework for hierarchical emotion distribution (ED) derived from textual cues. 
Our proposed pipeline supports three inference scenarios, as illustrated in Figure~\ref{fig:intro}: (a) Text-to-Speech (TTS) with Emotion Prediction, where the hierarchical emotion distribution (ED) is directly predicted from the input text; (b) TTS with Emotion Control, where the ED is predicted from the text and can be modified by users; and (c) Emotion Editing, where the ED is extracted from input audio and manually adjusted by users.
In ~\cite{ShoICASSP,ShoAPSIPA,ShoTAC},  a hierarchical ED was introduced to enable both global and fine-grained emotion modification in speech generation. Unlike prior single-step ED prediction approaches~\cite{ShoICASSP}, which treat different levels of emotion variance independently, we propose to explicitly model hierarchical dependencies by predicting EDs at the utterance, word, and phoneme levels in a multi-step manner. This structured approach ensures that higher-level emotional context influences lower-level prosodic details, resulting in a more coherent, expressive, and controllable emotional rendering. By leveraging multi-step ED prediction, our method provides fine-grained control, closely mimicking the way humans modulate speech—starting with an overall tone and refining intonation and articulation dynamically. This leads to an improved performance on both emotion expressiveness and speech naturalness. Furthermore, to demonstrate its flexibility, we explore two integration strategies for hierarchical ED: implementing it as a variance adaptor within FastSpeech2~\cite{ShoICASSP} and incorporating it as an external module compatible with any text-to-speech (TTS) model~\cite{ShoAPSIPA}. Through this approach, we bridge the gap between interpretability and fine-grained emotion control. Our contributions are summarized as follows\footnote{\textbf{Demo}: \url{https://shinshoji01.github.io/multi-step-prediction-HED/}}:
\vskip-0.8em
{
\setlength{\leftmargini}{15pt} 
\begin{itemize}
\setlength{\itemsep}{5pt}      %2. ブロック間の余白
\setlength{\parskip}{-5pt}      %4. 段落間余白．
\setlength{\itemindent}{0pt}   %5. 最初のインデント
\setlength{\labelsep}{5pt}     %6. item と文字の間

\item We introduce a multi-step prediction framework for hierarchical emotion distribution (ED), where the utterance-, word-, and phoneme-level EDs are derived from textual cues in successive steps. This structured approach ensures that higher-level emotional context guides low-level prosodic details, resulting in synthesized emotional speech that is both natural and expressive;

\item Leveraging the multi-step prediction of EDs, our method achieves refined control over emotion rendering and closely emulates human speech modulation. This approach not only enhances global and nuanced emotional expressiveness but also significantly improves performance in controlled emotional voices;
%offers an intuitive, quantitative method for multi-level emotion manipulation, enhancing emotion rendering and overall speech performance compared to our prior methods.

\item We explore two integration strategies for hierarchical ED into TTS systems: (1) embedding it as a variance adaptor within FastSpeech2 and (2) implementing it as an external module, making it adaptable to various text-to-speech (TTS) systems.

%\item Our approach enables users to analyze and manipulate emotion distributions at run-time, providing a flexible tool for adjusting emotion rendering. 

\end{itemize}
}

The rest of this paper is organized as follows: In Section 2, we discuss the related works. Section 3 describes our proposed methodology. In Section 4, we summarize experimental setups. In Section 5, we report our experiments and results. Section 6 concludes our study.

\section{Related Works}
In this section, we briefly introduce related studies to set the stage for our research and highlight the novelty of our contributions. We begin by discussing the hierarchical nature of speech emotions, emphasizing the need for multi-level and multi-step emotion modeling. We then review existing approaches to emotion rendering control in the TTS literature, identifying key advancements and gaps that our work addresses.

\subsection{Hierarchical Nature of Speech Emotion}

Speech emotions manifest hierarchically across three levels: utterance, word, and phoneme. At the utterance level, previous studies have shown that global prosodic patterns---such as pitch contour, range, mean, intonation, tempo, and rhythm---play a crucial role in conveying emotion~\cite{hy-utt}. At the word level, lexical cues shape emotional tone~\cite{hy-word1} and enhance intensity through emphasis~\cite{hy-word2}. Additionally, research suggests that when lexical and prosodic signals conflict, listeners tend to rely on prosodic cues to interpret emotion~\cite{hy-word3}. At the phoneme level, individual prosodic features such as pitch, energy, and duration contribute to emotional expression~\cite{hieprosody}, as supported by multiple studies~\cite{hy-ph1,hy-ph2,hy-ph3}. This hierarchical structure highlights the necessity of studying multi-level and multi-step emotion modeling for effective emotion modeling and control.

\subsection{Control of Speech Emotion}
%There has been a growing interest in enabling the control of synthesized emotions~\cite{emogst,emoscaler,PADcontrol}.
%Emotion-enhanced GST~\cite{emogst} incorporated an emotion recognition task to facilitate the modeling of emotion-related prosody. Some studies explored quantitative methods for controlling emotions through relative attributes~\cite{9003829,kun-intensity, kun-mix, pmsemotts}. Another study~\cite{interintra} delved into inter- and intra-class distances to achieve fine-grained control over recognizable intensity differences. EmoQ-TTS \cite{emoq} adopted a distance-based intensity quantization approach to capture emotion intensity.
%Unlike MsEmoTTS~\cite{msemotts}, which covers only utterance-level emotional change with fine-grained intensity changes, this work aims to offer quantitative and hierarchical control over emotions at various levels of speech units, e.g. phoneme, word, and utterance. It provides a versatile tool for editing emotion rendering in any speech segment.

Recent advancements in emotional TTS have significantly enhanced expressiveness; however, achieving interpretable emotion control remains a challenge. Prior studies have primarily focused on refining emotion intensity control by treating speech emotion as a global feature and manipulating representations or attributes derived from reference audio. For instance, \cite{oh2023semisupervised} enables utterance-level control via a speech mixer that predicts pseudo-labels and modulates features such as pitch, duration, and energy. \cite{zhang2023iemotts,li2022crossspeaker} further controlled emotion by manipulating speaker-disentangled representations in cross-speaker scenarios. Recent studies employ relative attributes \cite{parikh2011relative} for intensity control \cite{kun-intensity,9003829}.
%and classifier guidance in EmoDiff \cite{guo2023emodiff}. 
%Moreover, \cite{Schnell2021ImprovingET} uses an attention network with a sigmoid function, while \cite{luo2021emotion} utilizes a speech emotion recognizer and a prosody factor generator for prosodic modifications. 
Approaches for mixed emotions include manipulating relative attributes \cite{zhou2022speech}, incorporating noise mixing in diffusion models \cite{tang2023emomix}, and leveraging continuous emotional representations \cite{zhou2024emotional}. In contrast, EmoSphere-TTS \cite{cho24_interspeech,cho2024emosphere} models emotional complexity via a spherical emotion vector through Cartesian-spherical transformations, while \cite{jing2024enhancing} integrates computational paralinguistic text prompts to enhance emotional expressiveness.

To achieve fine-grained emotion control, several studies have explored segmental-level representations. For example, MsEmoTTS~\cite{msemotts} employs a global emotion label and modifies phoneme-level intensity via relative attributes. EmoQ-TTS~\cite{emoq} quantifies emotion intensity using a distance-based method, while the study in~\cite{interintra} refines control by examining inter- and intra-class distances. Additionally, CASEIN~\cite{cui2023casein} leverages a speech emotion recognition module to predict phoneme-level emotion distributions.  
\textcolor{black}{
These multi‐ or phoneme‐level approaches outperform utterance‐level modeling in emotional speech synthesis~\cite{msemotts,emoq,cui2023casein}
}
Building upon these efforts, our previous work introduced hierarchical emotion distribution (ED)\cite{ShoICASSP,ShoAPSIPA}, which enables multi-level emotion control, capturing both global and fine-grained emotional variations in speech synthesis. 

However, existing approaches to hierarchical emotion modeling still face several limitations. Previous methods \cite{ShoICASSP,ShoAPSIPA} rely on single-step prediction strategies, which treat different levels of emotion variance independently and fail to capture the contextual dependencies between hierarchical emotion distributions. Additionally, current techniques often lack a structured mechanism to ensure that higher-level emotions influence lower-level prosodic variations, leading to inconsistencies in emotion expressiveness. Furthermore, integration with TTS remains another challenge, as most approaches are either model-specific or require reference audio, limiting their adaptability across different TTS architectures. Addressing these gaps, we propose a multi-step ED prediction framework that models hierarchical ED progressively, ensuring a more interpretable, flexible, and effective approach to speech emotion rendering and control.

%\subsection{Speech Editing}
%Speech editing, particularly semantic editing, focuses on altering the textual information of speech while preserving the naturalness of the synthesized output~\cite{editts,contextaware,editspeech}. EditSpeech~\cite{editspeech} employed partial inference and bidirectional fusion techniques to effectively modify speech semantics. Another research, EdiTTS~\cite{editts}, utilized perturbations in the Gaussian space to allow for detailed audio edits, adjusting context and pitch, and enabling user-guided prosody editing. Yet another study~\cite{contextaware} presented a method for phoneme-level pitch adjustment and time-stretching, providing users with greater control over prosody.
%
%This work is motivated to provide a systematic approach for precisely controlling high-level prosodic patterns, which differs from the prior studies which primarily focused on modifying the physical speech attributes, e.g. pitch and duration.
%
%In this section, we delve into the details of the proposed fine-grained emotion editing mechanism. 
%We begin by formulating the problems and introducing a quantitative and hierarchical emotion control mechanism. We then detail the design of the hierarchical emotion distribution (ED) extractor, along with its associated training scheme. Lastly, we explain how to render the desired emotions by editing. 

\section{Multi-Step Prediction and Hierarchical Control of Emotion Intensity} \label{sec:methodology}

%\subsection{Problem Formulation}
%We consider two scenarios: ``Speech Editing'' and ``Text-to-Speech (TTS)". In the ``Speech Editing'' scenario, given an audio input and its transcript, we develop a model that enables emotion intensity control at the segmental level. In the ``TTS'' scenario, we rely solely on text input to predict an interpretable emotion representation that aligns with the textual content, thereby facilitating precise control over the emotion intensity of speech segments. 
We propose a novel approach, which supports the rendering of both single emotions and mixed emotions, that can be seamlessly integrated with various text-to-speech frameworks. Traditionally, speech databases label emotions at the utterance level, overlooking nuanced intensity variations within speech.  To address this challenge, we automatically generate fine-grained, quantitative intensity labels, which serve as ``soft labels'' for speech generation models, eliminating the need for manual annotation. This method effectively enhances emotion control, enables mixed-emotion rendering, and can be readily adapted to speech generation frameworks, including text-to-speech and voice conversion.

\subsection{Hierarchical Emotion Distribution (ED) Extractor} 

\begin{figure}[h!]
\centerline{\includegraphics[width=1.0\columnwidth]{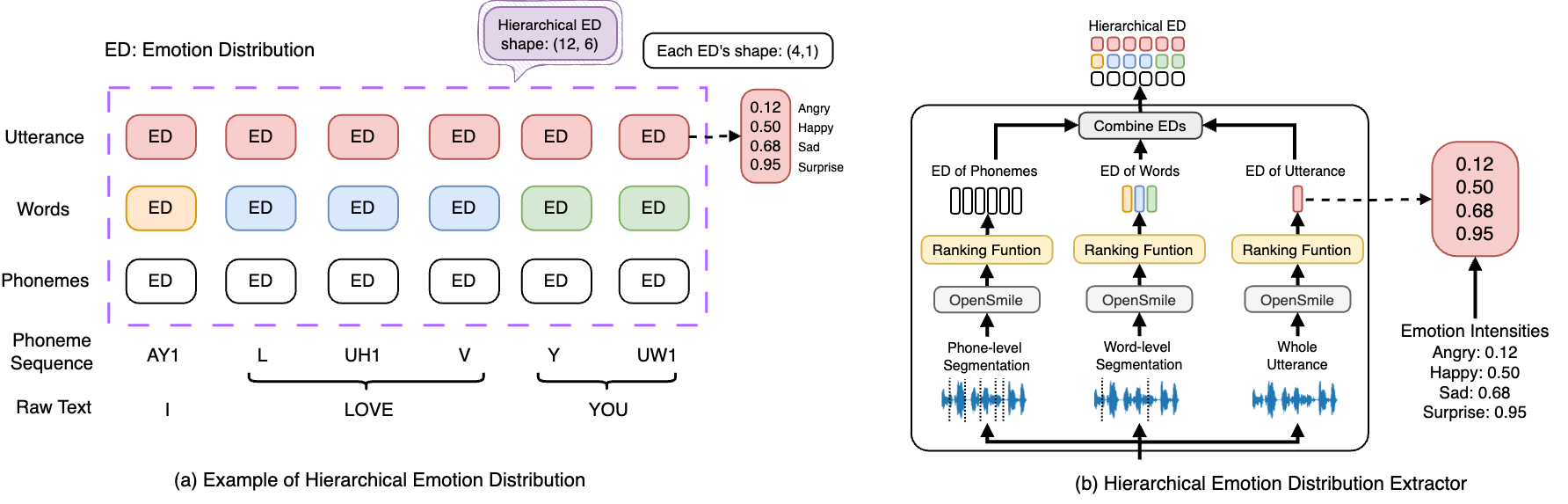}}
\caption{
(a) Example of Hierarchical Emotion Distribution (ED) including EDs at levels of utterance, words, and phonemes; (b) Diagram of Hierarchical ED Extractor.
}
\label{fig:hierarchical ED}
\end{figure}

Built upon our previous studies~\cite{ShoICASSP,ShoAPSIPA}, the hierarchical emotion distribution (ED) extraction module integrates OpenSmile feature extractors~\cite{opensmile} with pre-trained ranking functions at each segmental level to quantify emotion intensities in an utterance, as shown in Figure~\ref{fig:hierarchical ED}. Grounded in relative attributes~\cite{parikh2011relative}, our method measures emotion prominence by treating emotion style as a speech attribute and ranking its presence relative to other emotions, enabling a structured and interpretable approach to hierarchical emotion quantification.

Specifically, we define the ranking function as:
\begin{align}
f(\mathbf{x}_i) = \mathbf{w}^T \mathbf{x}_i + b
\end{align}
\noindent where $\mathbf{x}_i$, $\mathbf{w}$, and $b$ denote the acoustic features of the $i$-th sample, weight vector, and bias, respectively. We optimize these parameters using a support vector machine objective for binary classification (e.g., Angry vs. Non-angry)~\cite{svm} and normalize the outputs to the range $[0,1]$, with larger values indicating stronger emotion intensity. This process enables continuous labeling of training data and the quantification of emotion intensity in unseen utterances during run-time.

Figure~\ref{fig:hierarchical ED}(b) illustrates our hierarchical ED extractor. We begin by segmenting the input audio into phoneme, word, and utterance levels using the Montreal Forced Aligner~\cite{mfa}, and extracting an 88-dimensional feature set for each segment via OpenSMILE~\cite{opensmile}. The pre-trained ranking functions then estimate an ED vector for each segment, where each element represents the intensity of a specific emotion. To ensure hierarchical consistency, we duplicate the utterance-level ED across all phonemes and replicate the word-level ED for the corresponding phonemes, as shown in Figure~\ref{fig:hierarchical ED}(a). These hierarchical ED vectors are subsequently incorporated into TTS training, which will be introduced in the next subsection.
%\textcolor{black}{Finally, we integrate these hierarchical ED vectors with linguistic embeddings in the variance adaptor during training.}

\subsection{Multi-Step Modeling of Hierarchical ED for TTS}
We propose a multi-step strategy for modeling hierarchical emotion distribution (ED) to enable precise, multi-level control over emotion rendering in text-to-speech (TTS) synthesis. By progressively predicting ED at the utterance, word, and phoneme levels, our approach ensures that global emotional context guides local prosodic details. We explore two integration strategies to incorporate this multi-step hierarchical ED prediction into TTS frameworks.
%We explored two strategies for integrating hierarchical
%ED embedding into TTS frameworks: (1) External Integration (``External''): A model-agnostic approach applicable to any non-autoregressive TTS framework, incorporating hierarchical ED after text processing; (2) Variance Adaptor Integration (``VA''): A method that integrates hierarchical ED within the variance adaptor of FastSpeech2~\cite{fastspeech2}.

\subsubsection{External Integration}

\begin{figure}[h!]
\begin{center}
\centerline{\includegraphics[width=0.7\columnwidth]{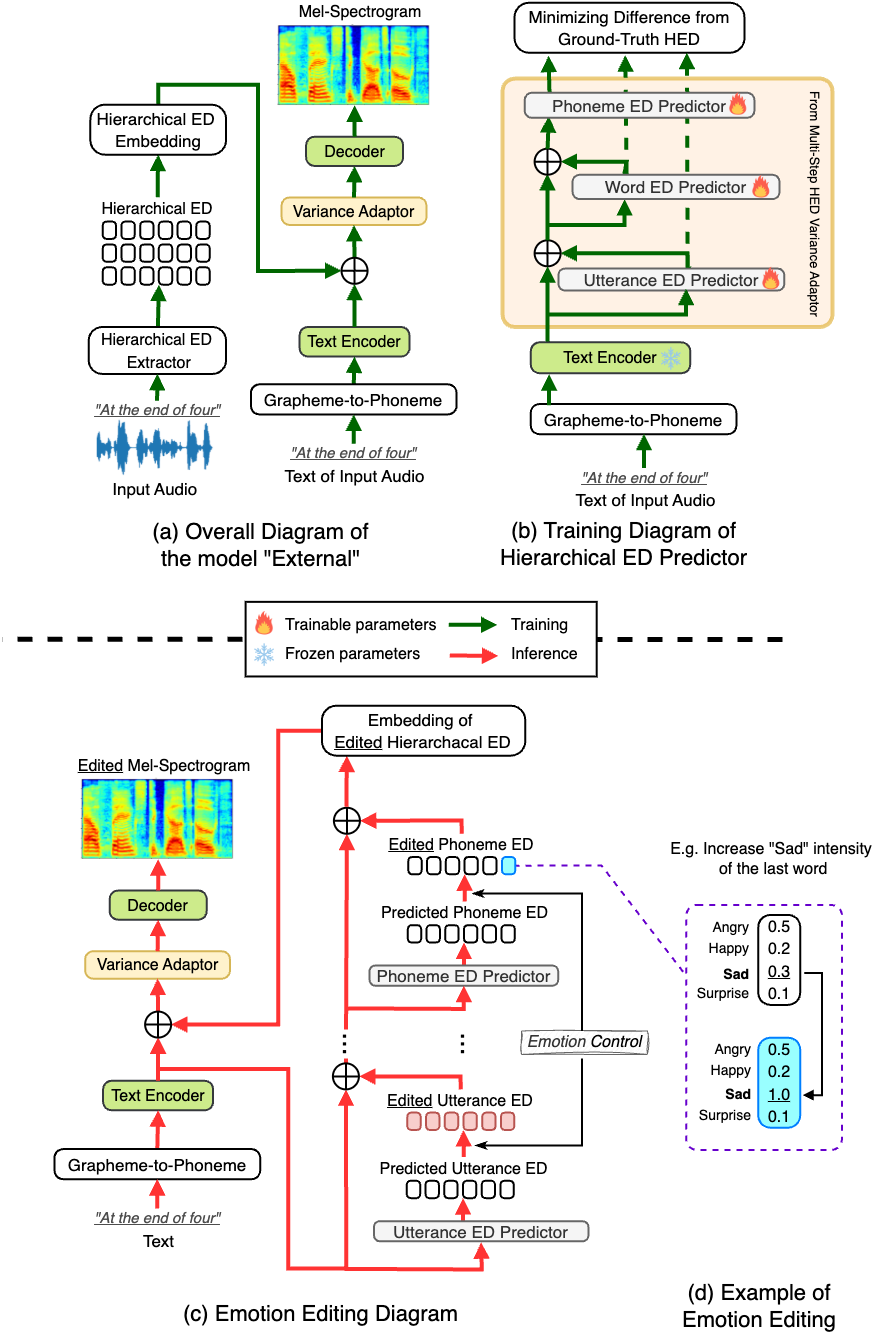}}
\end{center}
\caption{
Training and Inference Diagrams of the proposed framework using external integration (``External''): (a) Overall diagram; (b) Training diagram of hierarchical emotion distribution (ED) predictor; (c) Emotion editing (inference) diagram (d) Example of emotion editing.
}
\label{fig:external}
\end{figure}

%\noindent
%\textit{TTS Training Framework (``External''): }\\
In the external integration approach (``External''), as shown in Figure~\ref{fig:external}(a), we enhance a model-agnostic TTS pipeline by integrating a hierarchical ED embedding after text processing. In this paper, we choose FastSpeech2~\cite{fastspeech2} as our TTS backbone. A text encoder converts phoneme sequences into linguistic embeddings, while a fully connected network transforms the hierarchical ED into an ED embedding. A variance adaptor then predicts pitch, duration, and energy, followed by a decoder that reconstructs the Mel-spectrogram using an L1 loss. This design effectively captures emotion intensity and improves prosody. Since the ``External'' framework does not inherently predict hierarchical ED, we incorporate a dedicated multi-step hierarchical ED prediction module. In this module, EDs are predicted sequentially—starting from the utterance level, progressing to the word level, and finally to the phoneme level—while keeping the text encoder frozen as shown in Figure~\ref{fig:external}(b). This multi-step process allows a higher-level emotional context to guide finer, local prosodic adjustments.

%\noindent
%\\\textit{Prediction of Hierarchical ED from Textual Cues (``External''): }\\
%Since the ``External'' TTS model lacks inherent hierarchical ED prediction we trained ED predictors at the utterance, word, and phoneme levels in a multi-step hierarchical ED variance adaptor (Figure.\ref{fig:va}(b)) while freezing the text encoder, as shown in Figure.\ref{fig:external}(b). Consequently, our model predicts hierarchical EDs from textual cues sequentially from utterance, words, and phonemes.

\subsubsection{Variation Adaptor Integration}
\begin{figure}[h!]
\centerline{\includegraphics[width=0.9\columnwidth]{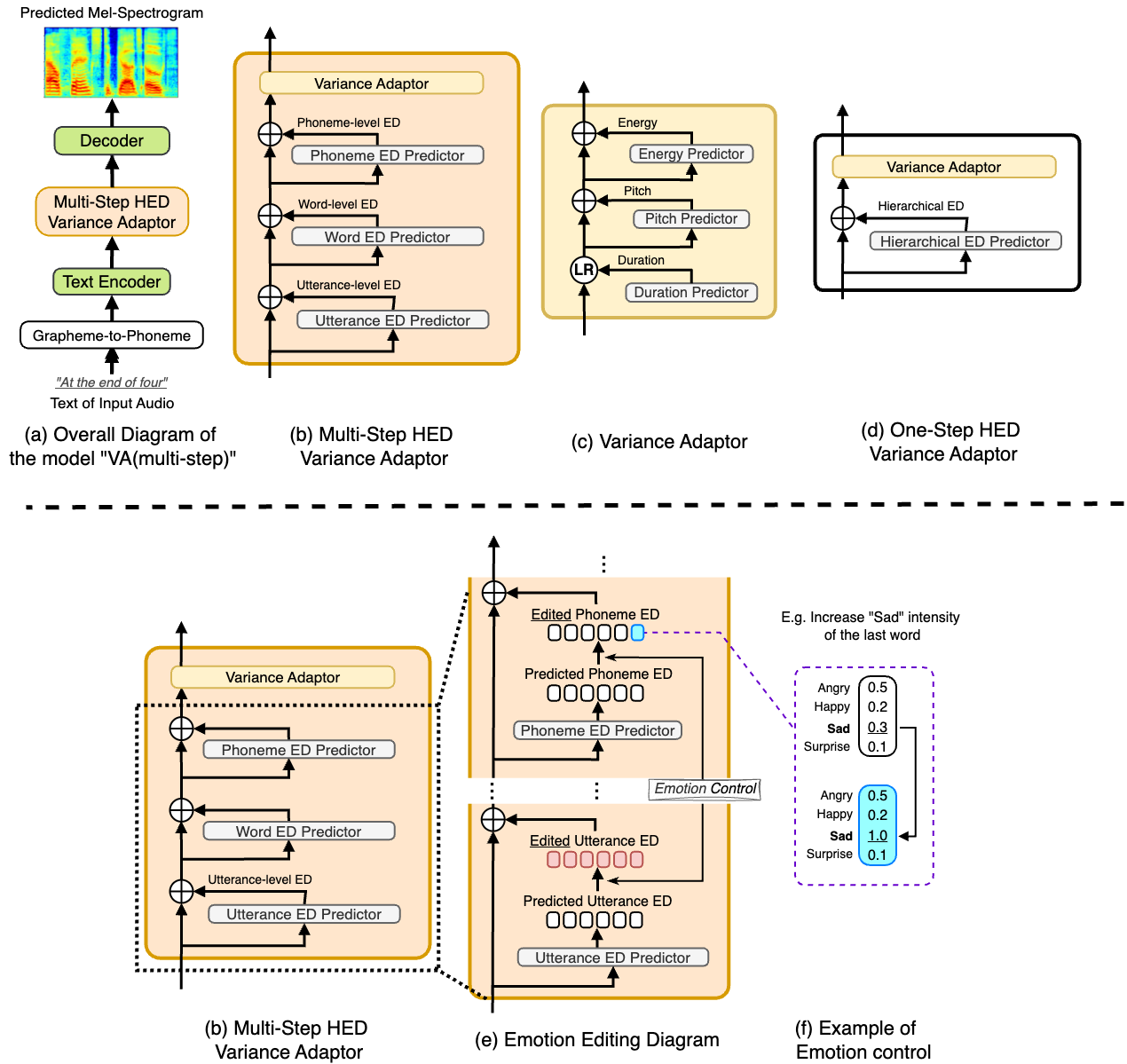}}
\caption{
Training and Inference Diagrams of the proposed framework using variance adaptor integration (``VA''): (a) Overall diagram; (b) Diagram of sequential hierarchical emotion distribution (hierarchical ED) variance adaptor; (c) Diagram of variance adaptor; (d) Diagram of parallel hierarchical ED variance adaptor (e) Emotion editing (inference) diagram (f) Example of emotion editing.
}
\label{fig:va}
\end{figure}

The variation adaptor integration approach (``VA'') integrates hierarchical ED modeling directly within the variance adaptor of FastSpeech2~\cite{fastspeech2}. This configuration extends the variance adaptor to jointly learn emotion representations and acoustic features, thereby tightly coupling emotion prediction with prosody generation. 
In contrast to the ``External'' setting, we train the linguistic encoder with a loss function that minimizes hierarchical ED differences. Specifically, we use a mean squared error (MSE) loss to reduce the discrepancy between the predicted and ground-truth EDs. 
Within the VA integration, ED prediction is also performed in a multi-step manner. The process begins with predicting the utterance-level ED to establish the global emotional tone, which then informs the word-level prediction. Finally, these outputs are combined to generate phoneme-level ED, enabling fine-grained emotion control, as shown in Figure~\ref{fig:va}(b). This hierarchical, multi-step approach ensures that a broader emotional context effectively influences lower-level prosodic details.

%\noindent
%\textit{TTS Training Framework (``VA''): }\\
%In the ``VA'' approach, we incorporated hierarchical ED modeling directly within the variance adaptor of FastSpeech2. In this configuration, we predict EDs sequentially from longer to shorter segments, denoted as ``Multi-Step hierarchical ED Variance Adaptor'' (Figure.\ref{fig:va}(b)). This approach jointly learns emotion representations and acoustic features.

%\noindent
%\\\textit{Prediction of Hierarchical ED from Textual Cues (``VA''): }\\
%Adopting the FastSpeech2 paradigm, we first predict the utterance-level ED using the variance adaptor, which subsequently informs the word-level prediction. We then combine these outputs to generate phoneme-level ED, enabling fine-grained emotion control during synthesis. 

\subsection{Run-time Emotion Editing and Control}
\label{sec: run-time}
During run-time, our framework supports two primary tasks: (1) Emotion Control and (2) Emotion Editing. For emotion control, where only text input is available, our model predicts a hierarchical emotion distribution that aligns with the textual content. This allows users to control the emotion intensity of individual speech segments, enabling fine-grained and quantifiable emotion rendering in synthesized speech. For the speech editing task, given an audio input and its corresponding transcript, our model extracts the hierarchical emotion distribution from the audio signal, enabling users control over emotion intensity for quantifiable emotion modification. In general, as depicted in Figure~\ref{fig:external}(c) and Figure~\ref{fig:va}(e), users are able to control the emotion rendering by adjusting the emotion distributions at three distinct levels, regardless of whether the ED is derived from audio or predicted from text.

%In the Speech Editing scenario, given an audio input and its transcript, our model derives the hierarchical emotion distribution (ED) from the audio signal to enable segment-level control over emotion intensity. 

%In the TTS scenario, where only text input is available, our model predicts an interpretable emotion representation that aligns with the textual content, thereby facilitating precise control over the emotion intensity of speech segments. As depicted in Figure.\ref{fig:external}(c) and Figure.\ref{fig:va}(e), users can edit the emotion rendering by adjusting the ED at three distinct levels, regardless of whether the ED is derived from audio or predicted from text.

\section{Experimental Setup}

%\subsection{Dataset} \label{sec:dataset}

We evaluated our system performance by conducting two experiments: (1) emotion prediction and (2) emotion editing. For emotion prediction, we train the TTS model on LibriTTS-R~\cite{LibriTTSR}, a multi-speaker dataset containing approximately 580 hours of recordings from 2,306 speakers. Specifically, we utilized the ``train-clean-100'' and ``train-clean-360'' subsets for model training. 

For the speech editing experiment, we used the Emotion Speech Dataset (ESD)~\cite{esd,zhou2021seen}, which comprises over 29 hours of emotional speech in five categories---Neutral, Angry, Happy, Sad, and Surprise---from 20 speakers (10 native English and 10 Mandarin). We exclusively used the English recordings and the full training split for TTS model training. We also use the ESD dataset to train the hierarchical ED extractor’s ranking functions, where we randomly selected 100 samples per speaker and emotion, resulting in a total of 5,000 audio samples for hierarchical ED extractor training.

\subsection{Model Architecture}
We choose FastSpeech2~\cite{fastspeech2} as our backbone, which comprises a text encoder, variance adaptor, and decoder. We use a transformer~\cite{transformer}-based encoder to convert input phoneme sequences into linguistic embeddings. Variance adaptors predict hierarchical ED, duration, pitch, and energy. We utilize a transformer-based decoder to synthesize mel-spectrograms from these features. Our loss function combines the L1 loss on mel-spectrograms with the mean squared error for prosodic predictions in the variance adaptor. To address multi-speaker scenarios, we integrate speaker embeddings from Resemblyzer~\cite{Resemblyzer} into the encoder output. We adopt the Adam optimizer~\cite{adam}. For TTS training, we use a batch size of 32 and perform 200,000 iterations over 48 hours on a single GPU. For text-based hierarchical ED prediction in ``External'', we conduct 100,000 iterations. The ED embedding layers consist of fully connected layers with a Tanh activation function. Finally, we employ HiFiGAN~\cite{hifigan} as the vocoder, trained on the ESD and LibriTTS-R datasets.

%We adopt the text-based emotion intensity predictor from MsEmoTTS~\cite{msemotts}, which uses two 1D convolution layers, layer normalization, and dropout.

\subsection{Baselines Comparison}
For emotion prediction experiments, we compared the model with our previous works~\cite{ShoICASSP,ShoAPSIPA} (``single-step''), which predict EDs from text in a single-step manner. For example, in the ``VA'' setting, our proposed model progressively predicted utterance-, word-, and phoneme-level EDs (Figure~\ref{fig:va}(b)), whereas the baseline~\cite{ShoICASSP} predicted all segments concurrently (Figure~\ref{fig:va}(d)). For emotion editing experiments, we re-implemented MsEmoTTS~\cite{msemotts} into the FastSpeech2 framework as the baseline (``MsEmoTTS'') to ensure a fair comparison.

\subsection{Evaluation Metrics}

For objective evaluation, we calculated the Word Error Rate (WER) using Whisper\footnote{Whisper Large: \url{https://github.com/openai/whisper}\label{whisper}}~\cite{whisper} to assess the system's robustness. To measure emotion similarity with the target, we evaluated spectral similarity using Mel-Cepstral Distortion (MCD)~\cite{mcd}, prosody alignment through pitch and energy distortion, and duration deviation using Frame Disturbance (FD)~\cite{fd}.

For subjective evaluation, we conducted three listening experiments with 20 participants, each evaluating 210 synthesized samples.
First, we conducted MUSHRA tests, where participants rated each speech sample on a scale from 0 to 100, with higher scores indicating better quality or greater similarity. The first test assessed speech naturalness, instructing participants to disregard noise and emotion. The second test evaluated emotion similarity, asking participants to rate the synthesized audio solely based on its emotional expressiveness while ignoring speech quality.
Additionally, we conducted best-worst scaling (BWS) tests~\cite{bws} to compare word-level emotion controllability between our model and the baseline. In BWStests, we adjusted the emotion intensity of three words per utterance to 0.0, 0.5, and 1.0, and evaluators selected the least and most expressive samples.

\section{Experiments and Results}
In this section, we present the experimental results for two tasks: (1) Emotion Prediction and (2) Emotion Editing. For Emotion Prediction, we evaluated our models based on speech quality and emotional expressiveness, supplemented by qualitative analysis. For Emotion Editing, we assessed the controllability of emotions, measuring how effectively the model modifies and adjusts emotion intensity for different speech segmental levels.

%We evaluated our system performance by conducting two experiments: text-based hierarchical emotion distribution (ED) prediction and speech editing. In the prediction experiment, we evaluated our models for speech quality and emotion expressiveness, complemented by qualitative analysis; in the latter, we assessed emotion controllability.

\subsection{Experiments with Emotion Prediction} 
%We performed text-based hierarchical ED prediction using LibriTTS-R. We evaluated both speech quality and emotion expressiveness as well as qualitative analysis.

%We compared synthesized audio samples under seven conditions, as detailed in Table~\ref{table:texthierarchical ED_quality}. The table categorizes the conditions into three columns: ``GT or Pred'', ``TTS Model'', and ``Pred Mode''. The column ``GT or Pred'' indicates whether we use ground-truth hierarchical ED (``GT'') or a text-predicted version. In the ``TTS Model'' column, ``VA'' and ``VA(Multi-Step)'' denote the TTS models employing the Single-Step hierarchical ED Variance Adaptor (Figure.\ref{fig:va}(d)) and the Multi-Step hierarchical ED Variance Adaptor (Figure.\ref{fig:va}(b)), respectively. Finally, the ``Pred Mode'' column specifies whether we predict the hierarchical ED sequentially from longer to shorter segments or in parallel for all segments.

We compared synthesized audio samples across seven different conditions, as detailed in Table~\ref{table:texthierarchical ED_quality} and Table~\ref{table:texthierarchical ED_expressiveness}. These tables organize these conditions into three key categories: ``GT or Pred'', ``TTS Model'', and ``Pred Mode'':

\begin{itemize}
    \item \textbf{GT or Pred}: This column specifies whether the hierarchical emotion distributions (EDs) are obtained from ground-truth data (GT) or predicted from text.
    \item \textbf{TTS Model}: This column indicates the TTS model used. “VA” refers to the model utilizing the Single-Step hierarchical ED Variance Adaptor (Figure~\ref{fig:va}(d)), while “VA (Multi-Step)” corresponds to the model employing the Multi-Step hierarchical ED Variance Adaptor (Figure~\ref{fig:va}(b)).
    \item \textbf{Pred Mode}: This column describes how the hierarchical ED is predicted. It is either generated progressively from longer to shorter segments (``Multi-Step'') or in parallel for all segments at once (``Single-Step'').
\end{itemize}
This configuration allows for a structured comparison of how different hierarchical ED configurations and prediction strategies impact synthesis quality and emotion expressiveness.

\subsubsection{Speech Quality Evaluation} 
%We conducted a MUSHRA test to assess speech naturalness, with higher scores indicating superior naturalness. We generated transcriptions from synthesized speech using Whisper\footref{whisper} and computed the word error rate (WER) against the ground-truth to evaluate intelligibility.

Table~\ref{table:texthierarchical ED_quality} summarizes the results of the MUSHRA and WER tests. With ground-truth hierarchical ED, the variance adapter with multi-step emotion modeling (“VA (Multi-Step)”) consistently outperforms the single-step approach (“VA”), achieving higher MUSHRA scores and lower WER. We also observe that, when using predicted hierarchical ED, the multi-step models significantly improve both speech naturalness and intelligibility compared to the single-step models. These findings suggest that aligning the EDs of shorter segments with those of longer segments is crucial for enhancing overall speech quality.

\begin{table}[!h]
\caption{Speech Quality Test Results: MUSHRA naturalness scores with 95\% confidence interval and Word Error Rate (WER). The column ``GT or Pred'' indicates whether we use ground-truth hierarchical ED (``GT'') or a text-predicted version. In the ``TTS Model'' column, ``VA'' and ``VA(Multi-Step)'' denote the TTS models employing the Single-Step hierarchical ED Variance Adaptor (Figure~\ref{fig:va}(d)) and the Multi-Step hierarchical ED Variance Adaptor (Figure~\ref{fig:va}(b)), respectively. Finally, the ``Pred Mode'' column specifies whether we predict the hierarchical ED sequentially from longer to shorter segments (Multi-Step) or in parallel for all segments (Single-Step). 
}
\label{table:texthierarchical ED_quality}
\centering
\scalebox{0.7}{
\begin{tabular}{ccccc}
\toprule
 \multicolumn{3}{c}{Hierarchical ED} & \multicolumn{2}{c}{Speech Quality}\\
\cmidrule(lr){1-3}\cmidrule(lr){4-5}
 GT or Pred & TTS Model & Pred Mode & MUSHRA ($\uparrow$) & WER ($\downarrow$)\\
\midrule
\multicolumn{3}{c}{--- Ground-Truth Speech Samples ---} & 79.4{\tiny $\pm$ 1.9 } & 2.16\\
\midrule
GT & External & - & 61.6{\tiny $\pm$ {2.2} } & 3.37\\
GT & VA & - & 57.5{\tiny $\pm$ {2.6} } & 3.11\\
GT & VA(Multi-Step) & - & \textbf{62.2}{\tiny $\pm$ {2.3} } & \textbf{2.48}\\
\midrule
Predicted & External & Single-Step & 50.7{\tiny $\pm$ {2.4} } & 3.80\\
Predicted & External & Multi-Step & \textbf{54.0}{\tiny $\pm$ {2.3} } & \textbf{3.25}\\
\addlinespace[0.1em]\hdashline\addlinespace[0.3em]
Predicted & VA & Single-Step & 52.2{\tiny $\pm$ {2.6} } & 4.61\\
Predicted & VA(Multi-Step) & Multi-Step & \textbf{53.2}{\tiny $\pm$ {2.4} } & \textbf{2.45}\\
\bottomrule
\end{tabular}
}
\end{table}

\subsubsection{Emotion Expressiveness Evaluation}
%We define emotion expressiveness as the similarity between the reference and synthesized audio. We conducted another MUSHRA test in which participants evaluated only the emotional similarity, ignoring speech quality. Additionally, we objectively measured emotion expressiveness using four metrics: Mel-Cepstral Distortion (MCD), Pitch/Energy Distortion, and Frame Disturbance (FD) to quantify duration deviation.

We further assess emotion expressiveness by conducting MUSHRA tests on emotional similarity and computing multiple objective prosody-related metrics, including Mel-Cepstral Distortion (MCD), Pitch/Energy Distortion, and Frame Disturbance (FD). As shown in Table~\ref{table:texthierarchical ED_expressiveness}, our proposed multi-step hierarchical ED prediction consistently achieves higher MUSHRA emotional similarity scores and lower distortion values across all objective metrics for both ground-truth and predicted hierarchical EDs. These results highlight the effectiveness of the multi-step scheme in enhancing emotion expressiveness and improving alignment with ground-truth emotions. However, we observe that in the ``VA'' setting, the multi-step scheme does not outperform the single-step approach in Pitch and FD. This discrepancy may stem from error accumulation across different levels of ED prediction (utterance, word, and phoneme levels). Compared to the ``External'' setting, where the linguistic encoder remains independent to emotion prediction, the ``VA'' setting incorporates hierarchical emotion distribution difference loss in training. This joint training may increase the model’s sensitivity to ED variations in longer segments, potentially leading to greater fluctuations in prosody-related metrics.

\begin{table}[!h]
\caption{Emotion Expressiveness Test Results with 95\% confidence interval: MUSHRA similarity scores, Mel-Cepstral Distortion (MCD), Pitch/Energy Distortion (Pitch/Energy), and Frame Disturbance (FD). The column ``GT or Pred'' indicates whether we use ground-truth hierarchical ED (``GT'') or a text-predicted version. In the ``TTS Model'' column, ``VA'' and ``VA(Multi-Step)'' denote the TTS models employing the Single-Step hierarchical ED Variance Adaptor (Figure~\ref{fig:va}(d)) and the Multi-Step hierarchical ED Variance Adaptor (Figure~\ref{fig:va}(b)), respectively. Finally, the ``Pred Mode'' column specifies whether we predict the hierarchical ED progressively from longer to shorter segments (Multi-Step) or in parallel for all segments (Single-Step).}
\label{table:texthierarchical ED_expressiveness}
\centering
\scalebox{0.63}{
\begin{tabular}{cccccccc}
\toprule
 \multicolumn{3}{c}{Hierarchical ED} & \multicolumn{5}{c}{Emotion Expressiveness}\\
\cmidrule(lr){1-3}\cmidrule(lr){4-8}
 GT or Pred & TTS Model & Pred Mode & MUSHRA ($\uparrow$) & MCD ($\downarrow$) & Pitch ($\downarrow$) & Energy ($\downarrow$) & FD ($\downarrow$)\\
\midrule
GT & External & - & \textbf{61.9}{\tiny $\pm$ {2.1} } & 5.88{\tiny $\pm$ {0.10} } & 15.6{\tiny $\pm$ {1.0} } & 0.363{\tiny $\pm$ {0.022} } & 24.8{\tiny $\pm$ {3.4} }\\
GT & VA & - & 55.8{\tiny $\pm$ {2.6} } & 6.48{\tiny $\pm$ {0.23} } & 16.1{\tiny $\pm$ {1.1} } & 0.386{\tiny $\pm$ {0.023} } & 25.5{\tiny $\pm$ {3.6} }\\
GT & VA(Multi-Step) & - & 61.9{\tiny $\pm$ {2.1} } & \textbf{5.62}{\tiny $\pm$ {0.11} } & \textbf{15.5}{\tiny $\pm$ {1.1} } & \textbf{0.348}{\tiny $\pm$ {0.020} } & \textbf{22.4}{\tiny $\pm$ {2.9} }\\
\midrule
Predicted & External & Single-Step & 47.2{\tiny $\pm$ {2.3} } & 7.59{\tiny $\pm$ {0.14} } & 18.2{\tiny $\pm$ {1.1} } & 0.438{\tiny $\pm$ {0.027} } & 46.3{\tiny $\pm$ {6.5} }\\
Predicted & External & Multi-Step & \textbf{51.9}{\tiny $\pm$ {2.2} } & \textbf{6.89}{\tiny $\pm$ {0.12} } & \textbf{16.9}{\tiny $\pm$ {1.2} } & \textbf{0.409}{\tiny $\pm$ {0.024} } & \textbf{42.6}{\tiny $\pm$ {4.7} }\\
\addlinespace[0.1em]\hdashline\addlinespace[0.3em]
Predicted & VA & Single-Step & 48.2{\tiny $\pm$ {2.5} } & 7.23{\tiny $\pm$ {0.20} } & \textbf{16.7}{\tiny $\pm$ {1.0} } & 0.426{\tiny $\pm$ {0.025} } & \textbf{41.4}{\tiny $\pm$ {4.7} }\\
Predicted & VA(Multi-Step) & Multi-Step & \textbf{49.1}{\tiny $\pm$ {2.2} } & \textbf{6.91}{\tiny $\pm$ {0.12} } & 17.2{\tiny $\pm$ {1.1} } & \textbf{0.416}{\tiny $\pm$ {0.025} } & 46.3{\tiny $\pm$ {5.1} }\\
\bottomrule
\end{tabular}
}
\end{table}

\subsubsection{Analysis on Hierarchical ED Prediction}

We further analyzed the predicted hierarchical emotion distribution (ED). Table~\ref{table:hierarchical ED_analysis} presents the mean absolute difference between the predicted and ground-truth values for each segment. 
The column ``Longer Segments'' denotes the longer segments used to predict shorter segments; ``GT'' indicates that ground-truth values were employed. For example, under the ``GT'' condition, we used the ground-truth utterance-level ED to predict the word-level ED, whereas under the ``Predicted'' condition, we utilized the predicted utterance-level ED.

Table~\ref{table:hierarchical ED_analysis} summarizes our results. We did not observe significant differences between single-step and multi-step predictions, despite substantial differences in synthesized audio performances (see Tables~\ref{table:texthierarchical ED_quality} and~\ref{table:texthierarchical ED_expressiveness}). These results suggest that our prediction modules not only reduce hierarchical ED differences but also generate emotion representations consistent with the textual emotional content in the audio domain. 
More importantly, when comparing ``Predicted'' and ``GT'' in the multi-step mode, we found error accumulation in ED prediction, evidenced by a smaller gap at the word level and an increased gap at the phoneme level. These findings suggest that our model prioritizes the dependency of EDs across segments over mere ED differences, which explains the improved speech naturalness (Table~\ref{table:texthierarchical ED_quality}) and only marginally better emotion expressiveness (Table~\ref{table:texthierarchical ED_expressiveness}).

\begin{table}[!h]
\caption{
Mean Absolute Difference of Hierarchical ED: differences between the predicted and the ground-truth hierarchical ED values. The column Longer ``Longer Segments'' denotes the longer segments used to predict shorter segments; ``GT'' indicates that ground-truth values were employed. For example, under the ``GT'' condition, we used the ground-truth utterance-level ED to predict the word-level ED, whereas under the ``Predicted'' condition, we utilized the predicted utterance-level ED.
}
\label{table:hierarchical ED_analysis}
\centering
\scalebox{0.7}{
\begin{tabular}{ccccccc}
\toprule
 \multicolumn{3}{c}{Hierarchical ED Condition} & \multicolumn{4}{c}{Hierarchical ED Difference}\\
\cmidrule(lr){1-3}\cmidrule(lr){4-7}
 TTS Model & Pred Mode & Longer Segments & Phonemes & Words & Utterance & Avg.\\
\midrule
External & Single-Step & - & 0.1333 & 0.1283 & 0.0594 & 0.1070\\
External & Multi-Step & Predicted & 0.1345 & 0.1297 & \textbf{0.0587} & 0.1077\\
External & Multi-Step & GT & \textbf{0.1214} & \textbf{0.1281} & 0.0587 & \textbf{0.1028}\\
\midrule
VA & Single-Step & - & 0.1358 & 0.1294 & \textbf{0.0599} & 0.1084\\
VA(Multi-Step) & Multi-Step & Predicted & 0.1356 & 0.1298 & 0.0601 & 0.1085\\
VA(Multi-Step) & Multi-Step & GT & \textbf{0.1230} & \textbf{0.1272} & 0.0601 & \textbf{0.1034}\\
\bottomrule
\end{tabular}
}
\end{table}

\begin{figure}
% \centerline{\includegraphics[width=0.8\columnwidth]{images/inoue3.png}}
\centerline{\includegraphics[width=0.8\columnwidth]{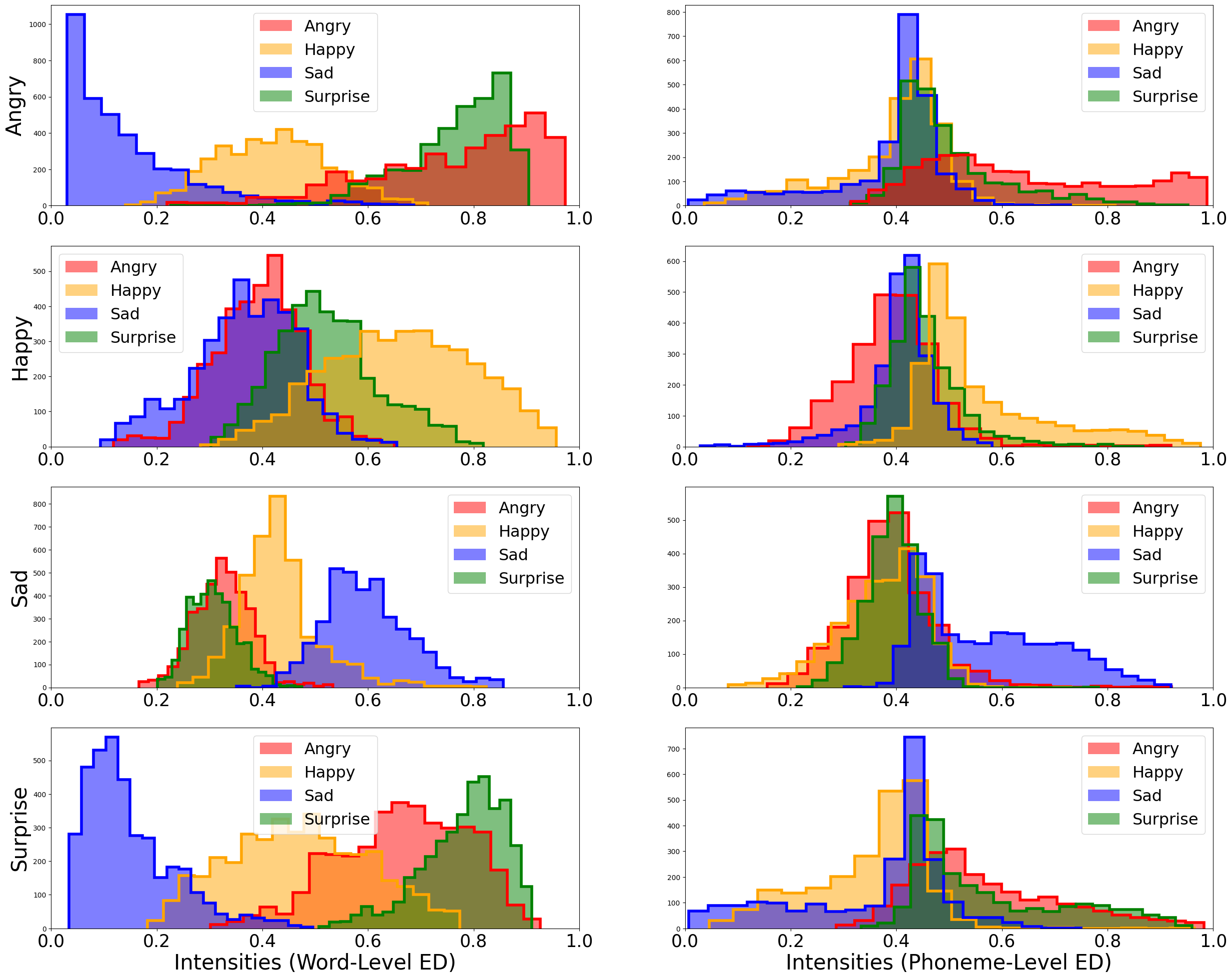}}
\caption{
Histograms of word-level and phoneme-level emotion distributions (EDs) for various intensified utterance-level EDs. Each row corresponds to an intensified emotion, and each column corresponds to a segment.
}
\label{fig:hierarchical ED_analysis}
\end{figure}

Next, we analyzed the predicted word- and phoneme-level EDs derived from different utterance-level EDs. Specifically, we systematically varied the intensity of a single emotion, setting it to 1.0 while keeping the intensities of all other emotions at 0.0. We then visualized the resulting ED distributions as histograms for each segment in Figure~\ref{fig:hierarchical ED_analysis}, where each row represents the intensified emotion and each column corresponds to a segment. 
We utilized a TTS model trained on the ESD dataset to highlight the impact of emotional variations in speech synthesis.
From Figure~\ref{fig:hierarchical ED_analysis}, we observe that both word- and phoneme-level EDs exhibit a positive correlation with their respective utterance-level ED values, with the correlation being stronger at the word level. This suggests that emotion propagation is more consistent across words than phonemes. Additionally, we find that word-level anger and surprise intensities display a notably stronger inter-correlation compared to other emotions, indicating that these two emotions may share similar prosodic and acoustic patterns at the word level. This observation aligns with psychological studies suggesting that anger and surprise often exhibit overlapping acoustic characteristics, such as increased pitch and energy.

\subsection{Experiments with Emotion Editing}

Following~\cite{ShoAPSIPA}, we evaluated our models' emotion controllability on the ESD dataset through subjective evaluations and objective analysis.
We first conducted best-worst scaling (BWS) tests~\cite{bws} to compare word-level emotion controllability between our model and the baseline MsEmoTTS~\cite{msemotts}. As presented in Table~\ref{table:control}, our model exhibited a stronger tendency than MsEmoTTS to select the least expressive sample at low intensity and the most expressive sample at high intensity across all five emotions. This trend was especially pronounced for Sad and Surprise emotions, where the distinction between intensity levels was more evident. These results demonstrate our model's ability to capture fine-grained variations in emotional intensity, ensuring more precise and consistent emotion rendering control compared to the baseline MsEmoTTS.
\begin{table}[h!]
\caption{
Best-Worst Scaling (BWS) Test Result: The value represents evaluator preferences (\%), with red and blue indicating the heatmap for the least expressive and most expressive audio, respectively.
}
\label{table:control}
\begin{center}
\scalebox{0.8}{
\begin{tabular}{m{0.6cm}c||m{0.4cm}m{0.4cm}m{0.4cm}m{0.4cm}|m{0.4cm}m{0.4cm}m{0.4cm}m{0.4cm}|}
& & \multicolumn{4}{c|}{Hierarchical ED} & \multicolumn{4}{c|}{MsEmoTTS}\\
& & Ang & Hap & Sad & Sur & Ang & Hap & Sad & Sur \\
\midrule
 & 0.0 & \cellcolor{red!91}{79} & \cellcolor{red!73}{63} & \cellcolor{red!77}{67} & \cellcolor{red!94}{81} & \cellcolor{red!48}{42} & \cellcolor{red!37}{32} & \cellcolor{red!24}{21} & \cellcolor{red!38}{33} \\
Least & 0.5 & \cellcolor{red!0}{0} & \cellcolor{red!32}{28} & \cellcolor{red!16}{14} & \cellcolor{red!16}{14} & \cellcolor{red!54}{47} & \cellcolor{red!54}{47} & \cellcolor{red!62}{54} & \cellcolor{red!48}{42} \\
 & 1.0 & \cellcolor{red!24}{21} & \cellcolor{red!10}{9} & \cellcolor{red!22}{19} & \cellcolor{red!5}{5} & \cellcolor{red!12}{11} & \cellcolor{red!24}{21} & \cellcolor{red!29}{25} & \cellcolor{red!29}{25} \\
\midrule
 & 0.0 & \cellcolor{cyan!12}{11} & \cellcolor{cyan!20}{18} & \cellcolor{cyan!18}{16} & \cellcolor{cyan!8}{7} & \cellcolor{cyan!12}{11} & \cellcolor{cyan!13}{12} & \cellcolor{cyan!34}{30} & \cellcolor{cyan!20}{18} \\
Most & 0.5 & \cellcolor{cyan!18}{16} & \cellcolor{cyan!8}{7} & \cellcolor{cyan!10}{9} & \cellcolor{cyan!8}{7} & \cellcolor{cyan!30}{26} & \cellcolor{cyan!18}{16} & \cellcolor{cyan!26}{23} & \cellcolor{cyan!34}{30} \\
 & 1.0 & \cellcolor{cyan!86}{74} & \cellcolor{cyan!87}{75} & \cellcolor{cyan!87}{75} & \cellcolor{cyan!100}{86} & \cellcolor{cyan!73}{63} & \cellcolor{cyan!83}{72} & \cellcolor{cyan!54}{47} & \cellcolor{cyan!61}{53} \\
\bottomrule
\end{tabular}
}
\end{center}
\end{table}

\begin{figure}
\centerline{\includegraphics[width=1.0\columnwidth]{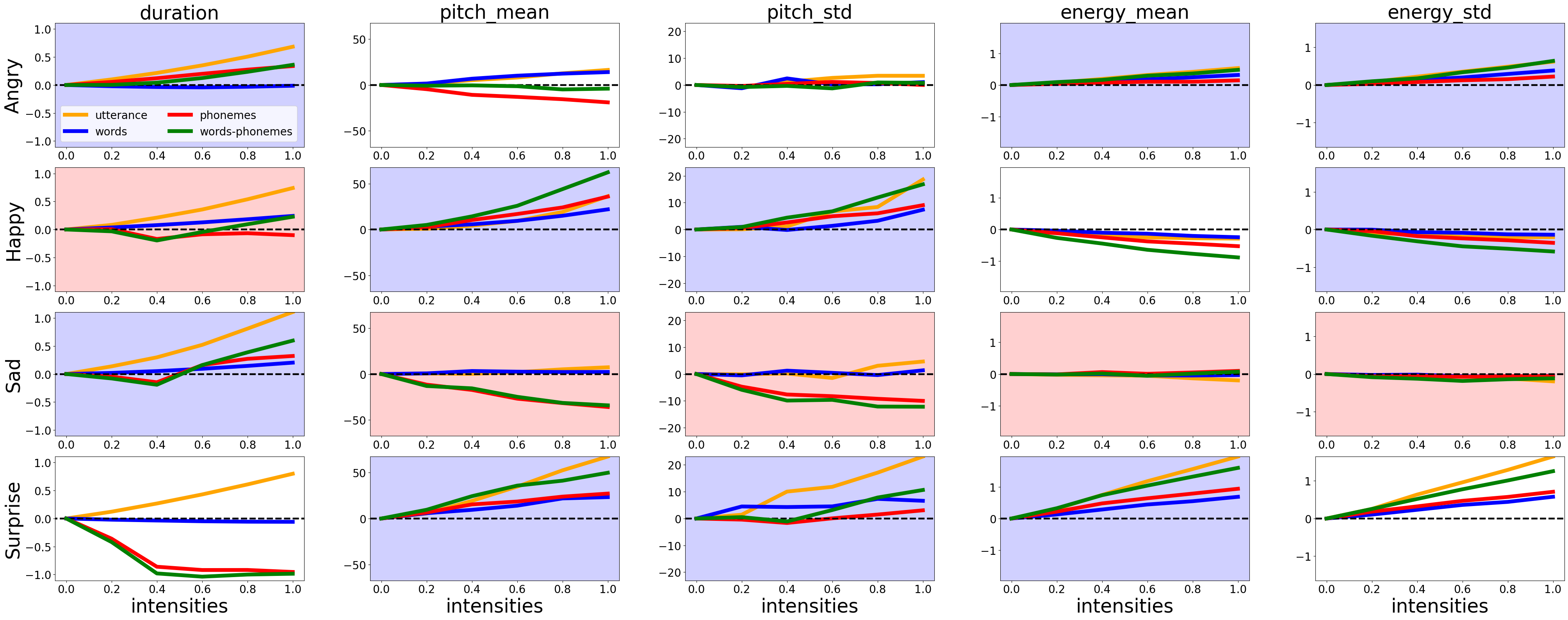}}
\caption{
The illustration of prosodic variants with intensity changes. The red background represents the expected negative trend, the blue indicates the expected positive trend, both summarized from the ESD dataset.
}
\label{fig:prosody_model}
\end{figure}

%\subsubsection{Objective Evaluations}
We further validated controllability across utterance, word, phoneme, and word-and-phoneme levels. We incremented emotion intensity from 0.0 to 1.0 and computed prosodic features such as duration and the mean/standard deviation of pitch and energy (Figure~\ref{fig:prosody_model}). Because duration values vary between levels, we standardized them prior to visualization. Prior literature~\cite{schuller2018speech} correlates these features with emotion intensity; for example, sadness is associated with a slower speaking rate and lower pitch and energy values. We analyzed the ESD dataset~\cite{esd} to examine these acoustic-emotion relationships. A red background indicates a negative trend, while blue is a positive trend with increasing intensity. 
Our model followed these expected trends, showing a positive correlation between happiness and mean pitch and a negative correlation between sadness and pitch. Additionally, editing both word and phoneme-level emotions produced significant prosodic changes, with the standard deviation of pitch at the utterance level aligning with our expectations.

\begin{figure}
\centerline{\includegraphics[width=0.95\columnwidth]{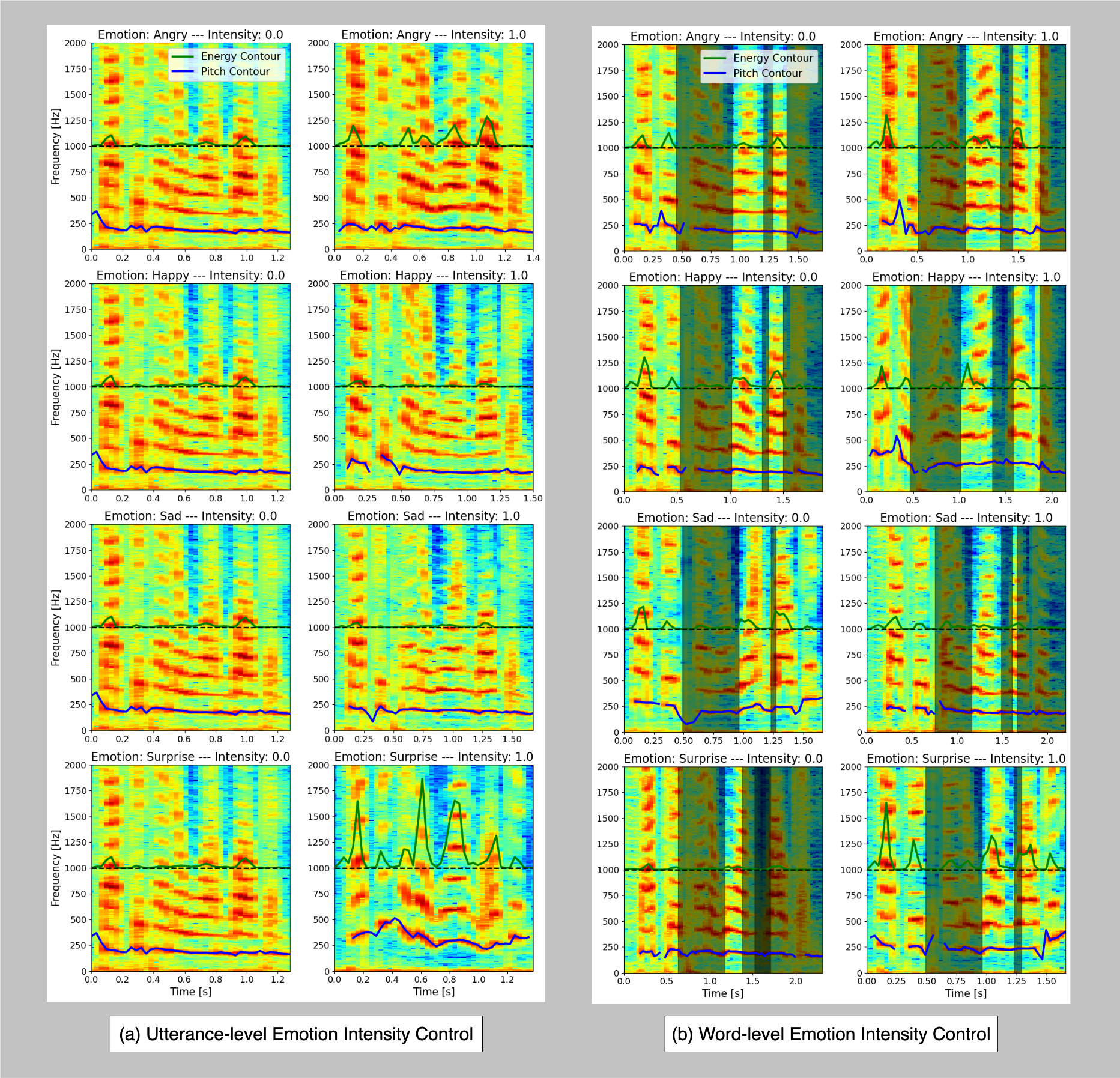}}
\caption{
Spectrograms of synthesized audio samples across different emotion intensities with pitch (blue) and energy (green) contours: the y-axis for energy contours are not relevant. (a) Utterance-level Emotion Intensity Control (b) Word-level Emotion Intensity Control.
}
\label{fig:qualitative}
\end{figure}

\textcolor{black}{
Figure~\ref{fig:qualitative} shows the spectrograms of synthesized audio samples with varying emotion intensities. We display pitch and energy contours (blue and green lines, respectively), noting that the y-axis for the energy contours is not relevant. Figures~\ref{fig:qualitative}(a) and (b) present utterance‐ and word‐level intensity control. Each row corresponds to a different emotion, with the first column depicting acoustic features at an emotion intensity of 0.0, and the second column at 1.0. In (b), the three highlighted areas indicate regions where the intensity has been modified. For both segments, for anger, we observe more pronounced energy spikes, at higher intensities. In happiness, pitch and energy patterns are similar, with higher pitch values at an intensity of 1.0. Sadness prolongs duration and stabilizes pitch contour as intensity increases. For surprise, we note a rise in both pitch contours and energy spikes. These results align with Figure~\ref{fig:prosody_model} and demonstrate that our model's ability to manipulate duration and pitch/energy according to emotion intensity.
}

\section{Conclusion}
We present a multi-step prediction framework for hierarchical emotion distribution (ED), enabling multi-level control of emotion rendering in speech synthesis. By modeling ED at the utterance, word, and phoneme levels progressively, our approach ensures that higher-level emotional context influences lower-level prosody, resulting in more natural and expressive speech. We integrate ED into TTS systems through two strategies: embedding it within the variance adaptor of FastSpeech2 and incorporating it as an external module for other non-autoregressive TTS models, making our method flexible and widely applicable. At runtime, users can quantitatively control emotion intensity, enhancing the interpretability and adaptability of emotional speech synthesis. Objective and subjective evaluations demonstrate that our approach significantly improves speech quality, expressiveness, and controllability. 
\textcolor{black}{In future work, we will extend our proposed method to additional languages, varied voice qualities, and more diverse emotional datasets, further exploring cross-lingual robustness and broadening the applicability of multi-level emotion intensity prediction.
}

%We introduce a system that sequentially predicts hierarchical emotion distributions (ED) from textual cues at the utterance, word, and phoneme levels. We integrate ED modeling with two strategies, rendering our approach applicable to FastSpeech2 and other non-autoregressive TTS models. At run-time, users can quantitatively control emotion rendering. Objective and subjective evaluations confirm our method's efficacy in enhancing speech quality, emotion expressiveness, and controllability.

%BACKMATTER SEE DOCUMENTATION
%\nocite{*}
\printbibliography

@misc{zhou2023mixedevc,
      title={Mixed-EVC: Mixed Emotion Synthesis and Control in Voice Conversion}, 
      author={Kun Zhou and Berrak Sisman and Carlos Busso and Bin Ma and Haizhou Li},
      year={2023},
      eprint={2210.13756},
      archivePrefix={arXiv},
      primaryClass={eess.AS}
}

@article{triantafyllopoulos2024expressivity,
  title={Expressivity and Speech Synthesis},
  author={Triantafyllopoulos, Andreas and Schuller, Bj{\"o}rn W},
  journal={arXiv preprint arXiv:2404.19363},
  year={2024}
}

@inproceedings{ming2016deep,
  title={Deep Bidirectional LSTM Modeling of Timbre and Prosody for Emotional Voice Conversion.},
  author={Ming, Huaiping and Huang, Dong-Yan and Xie, Lei and Wu, Jie and Dong, Minghui and Li, Haizhou},
  booktitle={Interspeech},
  pages={2453--2457},
  year={2016}
}

@article{esd,
  title={Emotional voice conversion: Theory, databases and ESD},
  author={Zhou, Kun and Sisman, Berrak and Liu, Rui and Li, Haizhou},
  journal={Speech Communication},
  volume={137},
  pages={1--18},
  year={2022},
  publisher={Elsevier}
}

@article{hirschberg2006pragmatics,
  title={Pragmatics and intonation},
  author={Hirschberg, Julia},
  journal={The handbook of pragmatics},
  pages={515--537},
  year={2006},
  publisher={Wiley Online Library}
}

@inproceedings{zhou2021seen,
  title={Seen and unseen emotional style transfer for voice conversion with a new emotional speech dataset},
  author={Zhou, Kun and Sisman, Berrak and Liu, Rui and Li, Haizhou},
  booktitle={ICASSP 2021-2021 IEEE International Conference on Acoustics, Speech and Signal Processing (ICASSP)},
  pages={920--924},
  year={2021},
  organization={IEEE}
}

@inproceedings{zhou2020transforming,
  title={Transforming Spectrum and Prosody for Emotional Voice Conversion with Non-Parallel Training Data},
  author={Zhou, Kun and Sisman, Berrak and Li, Haizhou},
  booktitle={Proc. Odyssey 2020 The Speaker and Language Recognition Workshop},
  pages={230--237},
  year={2020}
}

@Inbook{hieprosody,
author="Krothapalli, Sreenivasa Rao
and Koolagudi, Shashidhar G.",
title="Emotion Recognition Using Prosodic Information",
bookTitle="Emotion Recognition using Speech Features",
year="2013",
publisher="Springer New York",
address="New York, NY",
pages="79--91",
isbn="978-1-4614-5143-3",
doi="10.1007/978-1-4614-5143-3_5",
url="https://doi.org/10.1007/978-1-4614-5143-3_5"
}

@misc{adam,
      title={Adam: A Method for Stochastic Optimization}, 
      author={Diederik P. Kingma and Jimmy Ba},
      year={2017},
      eprint={1412.6980},
      archivePrefix={arXiv},
      primaryClass={cs.LG}
}

@inproceedings{parikh2011relative,
  title={Relative attributes},
  author={Parikh, Devi and Grauman, Kristen},
  booktitle={2011 International Conference on Computer Vision},
  pages={503--510},
  year={2011},
  organization={IEEE}
}

@misc{transformer,
      title={Attention Is All You Need}, 
      author={Ashish Vaswani and Noam Shazeer and Niki Parmar and Jakob Uszkoreit and Llion Jones and Aidan N. Gomez and Lukasz Kaiser and Illia Polosukhin},
      year={2017},
      eprint={1706.03762},
      archivePrefix={arXiv},
      primaryClass={cs.CL}
}

@INPROCEEDINGS{mcd,
  author={Kubichek, R.},
  booktitle={Proceedings of IEEE Pacific Rim Conference on Communications Computers and Signal Processing}, 
  title={Mel-cepstral distance measure for objective speech quality assessment}, 
  year={1993},
  volume={1},
  number={},
  pages={125-128 vol.1},
  doi={10.1109/PACRIM.1993.407206}}

@article{kun2022emotion,
  title={Emotion modelling for speech generation},
  author={KUN, ZHOU},
  year={2022}
}

@article{schuller2018speech,
  title={Speech emotion recognition: Two decades in a nutshell, benchmarks, and ongoing trends},
  author={Schuller, Bj{\"o}rn W},
  journal={Communications of the ACM},
  volume={61},
  number={5},
  pages={90--99},
  year={2018},
  publisher={ACM New York, NY, USA}
}

@article{el2011survey,
  title={Survey on speech emotion recognition: Features, classification schemes, and databases},
  author={El Ayadi, Moataz and Kamel, Mohamed S and Karray, Fakhri},
  journal={Pattern recognition},
  volume={44},
  number={3},
  pages={572--587},
  year={2011},
  publisher={Elsevier}
}

@INPROCEEDINGS{fd,
  author={Sisman, Berrak and Lee, Grandee and Li, Haizhou and Tan, Kay Chen},
  booktitle={2017 International Conference on Asian Language Processing (IALP)}, 
  title={On the analysis and evaluation of prosody conversion techniques}, 
  year={2017},
  volume={},
  number={},
  pages={44-47},
  doi={10.1109/IALP.2017.8300542}}

@INPROCEEDINGS{9003829,
  author={Zhu, Xiaolian and Yang, Shan and Yang, Geng and Xie, Lei},
  booktitle={2019 IEEE Automatic Speech Recognition and Understanding Workshop (ASRU)}, 
  title={Controlling Emotion Strength with Relative Attribute for End-to-End Speech Synthesis}, 
  year={2019},
  volume={},
  number={},
  pages={192-199},
  doi={10.1109/ASRU46091.2019.9003829}}

@inproceedings{bws,
    title = "Best-Worst Scaling More Reliable than Rating Scales: A Case Study on Sentiment Intensity Annotation",
    author = "Kiritchenko, Svetlana  and
      Mohammad, Saif",
    editor = "Barzilay, Regina  and
      Kan, Min-Yen",
    booktitle = "Proceedings of the 55th Annual Meeting of the Association for Computational Linguistics (Volume 2: Short Papers)",
    month = jul,
    year = "2017",
    address = "Vancouver, Canada",
    publisher = "Association for Computational Linguistics",
    url = "https://aclanthology.org/P17-2074",
    doi = "10.18653/v1/P17-2074",
    pages = "465--470",
}

@misc{lei2022multiscale,
      title={Towards Multi-Scale Speaking Style Modelling with Hierarchical Context Information for Mandarin Speech Synthesis}, 
      author={Shun Lei and Yixuan Zhou and Liyang Chen and Jiankun Hu and Zhiyong Wu and Shiyin Kang and Helen Meng},
      year={2022},
      eprint={2204.02743},
      archivePrefix={arXiv},
      primaryClass={cs.SD}
}

@article{hy-utt,
title = {Intonation and Emotion: Influence of Pitch Levels and Contour Type on Creating Emotions},
journal = {Journal of Voice},
volume = {25},
number = {1},
pages = {e25-e34},
year = {2011},
issn = {0892-1997},
doi = {https://doi.org/10.1016/j.jvoice.2010.02.002},
url = {https://www.sciencedirect.com/science/article/pii/S0892199710000378},
author = {Emma Rodero},
}

@article{hy-word1,
author = {Warriner, Amy and Kuperman, Victor and Brysbaert, Marc},
year = {2013},
month = {02},
pages = {},
title = {Norms of valence, arousal, and dominance for 13,915 English lemmas},
volume = {45},
journal = {Behavior research methods},
doi = {10.3758/s13428-012-0314-x}
}

@article{hy-word2,
title = {The influence of pitch range, duration, amplitude and spectral features on the interpretation of the rise-fall-rise intonation contour in English},
journal = {Journal of Phonetics},
volume = {20},
number = {2},
pages = {241-251},
year = {1992},
issn = {0095-4470},
doi = {https://doi.org/10.1016/S0095-4470(19)30625-4},
url = {https://www.sciencedirect.com/science/article/pii/S0095447019306254},
author = {Julia Hirschberg and Gregory Ward},
}

@article{hy-word3,
author = {Snedeker, Jesse and Trueswell, John},
year = {2003},
month = {01},
pages = {103-130},
title = {Using prosody to avoid ambiguity: Effects of speaker awareness and referential context},
volume = {48},
journal = {Journal of Memory and Language},
doi = {10.1016/S0749-596X(02)00519-3}
}

@inproceedings{opensmile,
author = {Eyben, Florian and Wöllmer, Martin and Schuller, Björn},
year = {2010},
month = {01},
pages = {1459-1462},
title = {openSMILE -- The Munich Versatile and Fast Open-Source Audio Feature Extractor},
journal = {MM'10 - Proceedings of the ACM Multimedia 2010 International Conference},
doi = {10.1145/1873951.1874246}
}

@ARTICLE{hy-ph1,
  author={Busso, Carlos and Lee, Sungbok and Narayanan, Shrikanth},
  journal={IEEE Transactions on Audio, Speech, and Language Processing}, 
  title={Analysis of Emotionally Salient Aspects of Fundamental Frequency for Emotion Detection}, 
  year={2009},
  volume={17},
  number={4},
  pages={582-596},
  doi={10.1109/TASL.2008.2009578}}

@article{hy-ph2,
title = {Survey on speech emotion recognition: Features, classification schemes, and databases},
journal = {Pattern Recognition},
volume = {44},
number = {3},
pages = {572-587},
year = {2011},
issn = {0031-3203},
doi = {https://doi.org/10.1016/j.patcog.2010.09.020},
url = {https://www.sciencedirect.com/science/article/pii/S0031320310004619},
author = {Moataz {El Ayadi} and Mohamed S. Kamel and Fakhri Karray},
}

@article{hy-ph3,
  author = {Leinonen, Lasse and Hiltunen, Tuija and Linnankoski, Ilkka and Laakso, Minna-Liisa},
  title = {Expression of emotional--motivational connotations with a one-word utterance},
  journal = {Journal of the Acoustical Society of America},
  volume = {102},
  number = {3},
  pages = {1853--1863},
  year = {1997},
  url = {https://doi.org/10.1121/1.420109}
}

@misc{fastspeech2,
      title={FastSpeech 2: Fast and High-Quality End-to-End Text to Speech}, 
      author={Yi Ren and Chenxu Hu and Xu Tan and Tao Qin and Sheng Zhao and Zhou Zhao and Tie-Yan Liu},
      year={2022},
      eprint={2006.04558},
      archivePrefix={arXiv},
      primaryClass={eess.AS}
}

@ARTICLE{kun-intensity,
  author={Zhou, Kun and Sisman, Berrak and Rana, Rajib and Schuller, Björn W. and Li, Haizhou},
  journal={IEEE Transactions on Affective Computing}, 
  title={Emotion Intensity and its Control for Emotional Voice Conversion}, 
  year={2023},
  volume={14},
  number={1},
  pages={31-48},
  doi={10.1109/TAFFC.2022.3175578}}

@misc{msemotts,
      title={MsEmoTTS: Multi-scale emotion transfer, prediction, and control for emotional speech synthesis}, 
      author={Yi Lei and Shan Yang and Xinsheng Wang and Lei Xie},
      year={2022},
      eprint={2201.06460},
      archivePrefix={arXiv},
      primaryClass={cs.SD}
}

@INPROCEEDINGS{emoq,
  author={Im, Chae-Bin and Lee, Sang-Hoon and Kim, Seung-Bin and Lee, Seong-Whan},
  booktitle={ICASSP 2022 - 2022 IEEE International Conference on Acoustics, Speech and Signal Processing (ICASSP)}, 
  title={EMOQ-TTS: Emotion Intensity Quantization for Fine-Grained Controllable Emotional Text-to-Speech}, 
  year={2022},
  volume={},
  number={},
  pages={6317-6321},
  doi={10.1109/ICASSP43922.2022.9747098}}

@INPROCEEDINGS{interintra,
  author={Wang, Shijun and Guonason, Jon and Borth, Damian},
  booktitle={ICASSP 2023 - 2023 IEEE International Conference on Acoustics, Speech and Signal Processing (ICASSP)}, 
  title={Fine-Grained Emotional Control of Text-to-Speech: Learning to Rank Inter- and Intra-Class Emotion Intensities}, 
  year={2023},
  volume={},
  number={},
  pages={1-5},
  doi={10.1109/ICASSP49357.2023.10097118}}

@inproceedings{mfa,
  author={McAuliffe, Michael and Socolof, Michaela and Mihuc, Sarah and Wagner, Michael and Sonderegger, Morgan},
  title={{Montreal Forced Aligner: Trainable Text-Speech Alignment Using Kaldi}},
  year=2017,
  booktitle={Proc. Interspeech 2017},
  pages={498--502},
  doi={10.21437/Interspeech.2017-1386}
}

@article{svm,
  title={Support-vector networks},
  author={Cortes, Corinna and Vapnik, Vladimir},
  journal={Machine learning},
  volume={20},
  number={3},
  pages={273--297},
  year={1995},
  publisher={Springer}
}

@misc{hifigan,
      title={HiFi-GAN: Generative Adversarial Networks for Efficient and High Fidelity Speech Synthesis}, 
      author={Jungil Kong and Jaehyeon Kim and Jaekyoung Bae},
      year={2020},
      eprint={2010.05646},
      archivePrefix={arXiv},
      primaryClass={cs.SD}
}

@article{triantafyllopoulos2023overview,
  title={An overview of affective speech synthesis and conversion in the deep learning era},
  author={Triantafyllopoulos, Andreas and Schuller, Bj{\"o}rn W and {\.I}ymen, G{\"o}k{\c{c}}e and Sezgin, Metin and He, Xiangheng and Yang, Zijiang and Tzirakis, Panagiotis and Liu, Shuo and Mertes, Silvan and Andr{\'e}, Elisabeth and others},
  journal={Proceedings of the IEEE},
  year={2023},
  publisher={IEEE}
}

@article{xu2011speech,
  title={Speech prosody: A methodological review},
  author={Xu, Yi},
  journal={Journal of Speech Sciences},
  volume={1},
  number={1},
  pages={85--115},
  year={2011}
}

@inproceedings{latorre2008multilevel,
  title={Multilevel parametric-base F0 model for speech synthesis},
  author={Latorre, Javier and Akamine, Masami},
  booktitle={Ninth Annual Conference of the International Speech Communication Association},
  year={2008}}

@INPROCEEDINGS{ShoICASSP,
  author={Inoue, Sho and Zhou, Kun and Wang, Shuai and Li, Haizhou},
  booktitle={ICASSP 2024 - 2024 IEEE International Conference on Acoustics, Speech and Signal Processing (ICASSP)}, 
  title={Hierarchical Emotion Prediction and Control in Text-to-Speech Synthesis}, 
  year={2024},
  volume={},
  number={},
  pages={10601-10605},
  doi={10.1109/ICASSP48485.2024.10445996}}

@misc{ShoTAC,
      title={Hierarchical Control of Emotion Rendering in Speech Synthesis}, 
      author={Sho Inoue and Kun Zhou and Shuai Wang and Haizhou Li},
      year={2025},
      eprint={2412.12498},
      archivePrefix={arXiv},
      primaryClass={cs.SD},
      url={https://arxiv.org/abs/2412.12498}, 
}

@article{ShoAPSIPA,
  title={Fine-Grained Quantitative Emotion Editing for Speech Generation},
  author={Sho Inoue and Kun Zhou and Shuai Wang and Haizhou Li},
  journal={2024 Asia Pacific Signal and Information Processing Association Annual Summit and Conference (APSIPA ASC)},
  year={2024},
  pages={1-6},
  url={https://api.semanticscholar.org/CorpusID:268248771}
}

@article{LibriTTSR,
  title={Libritts-r: A restored multi-speaker text-to-speech corpus},
  author={Koizumi, Yuma and Zen, Heiga and Karita, Shigeki and Ding, Yifan and Yatabe, Kohei and Morioka, Nobuyuki and Bacchiani, Michiel and Zhang, Yu and Han, Wei and Bapna, Ankur},
  journal={arXiv preprint arXiv:2305.18802},
  year={2023}
}

@misc{Resemblyzer,
      title={Generalized End-to-End Loss for Speaker Verification}, 
      author={Li Wan and Quan Wang and Alan Papir and Ignacio Lopez Moreno},
      year={2020},
      eprint={1710.10467},
      archivePrefix={arXiv},
      primaryClass={eess.AS}
}

@misc{whisper,
      title={Robust Speech Recognition via Large-Scale Weak Supervision}, 
      author={Alec Radford and Jong Wook Kim and Tao Xu and Greg Brockman and Christine McLeavey and Ilya Sutskever},
      year={2022},
      eprint={2212.04356},
      archivePrefix={arXiv},
      primaryClass={eess.AS},
      url={https://arxiv.org/abs/2212.04356}, 
}

@misc{oh2023semisupervised,
      title={Semi-supervised learning for continuous emotional intensity controllable speech synthesis with disentangled representations}, 
      author={Yoori Oh and Juheon Lee and Yoseob Han and Kyogu Lee},
      year={2023},
      eprint={2211.06160},
      archivePrefix={arXiv},
      primaryClass={eess.AS}
}

@misc{zhang2023iemotts,
      title={iEmoTTS: Toward Robust Cross-Speaker Emotion Transfer and Control for Speech Synthesis based on Disentanglement between Prosody and Timbre}, 
      author={Guangyan Zhang and Ying Qin and Wenjie Zhang and Jialun Wu and Mei Li and Yutao Gai and Feijun Jiang and Tan Lee},
      year={2023},
      eprint={2206.14866},
      archivePrefix={arXiv},
      primaryClass={eess.AS}
}

@misc{li2022crossspeaker,
      title={Cross-speaker emotion disentangling and transfer for end-to-end speech synthesis}, 
      author={Tao Li and Xinsheng Wang and Qicong Xie and Zhichao Wang and Lei Xie},
      year={2022},
      eprint={2109.06733},
      archivePrefix={arXiv},
      primaryClass={cs.SD}
}

@misc{zhou2022speech,
      title={Speech Synthesis with Mixed Emotions}, 
      author={Kun Zhou and Berrak Sisman and Rajib Rana and B. W. Schuller and Haizhou Li},
      year={2022},
      eprint={2208.05890},
      archivePrefix={arXiv},
      primaryClass={cs.CL}
}

@misc{tang2023emomix,
      title={EmoMix: Emotion Mixing via Diffusion Models for Emotional Speech Synthesis}, 
      author={Haobin Tang and Xulong Zhang and Jianzong Wang and Ning Cheng and Jing Xiao},
      year={2023},
      eprint={2306.00648},
      archivePrefix={arXiv},
      primaryClass={cs.SD}
}

@misc{cui2023casein,
      title={CASEIN: Cascading Explicit and Implicit Control for Fine-grained Emotion Intensity Regulation}, 
      author={Yuhao Cui and Xiongwei Wang and Zhongzhou Zhao and Wei Zhou and Haiqing Chen},
      year={2023},
      eprint={2307.00020},
      archivePrefix={arXiv},
      primaryClass={cs.SD}
}

@inproceedings{cho24_interspeech,
  title     = {EmoSphere-TTS: Emotional Style and Intensity Modeling via Spherical Emotion Vector for Controllable Emotional Text-to-Speech},
  author    = {Deok-Hyeon Cho and Hyung-Seok Oh and Seung-Bin Kim and Sang-Hoon Lee and Seong-Whan Lee},
  year      = {2024},
  booktitle = {Interspeech 2024},
  pages     = {1810--1814},
  doi       = {10.21437/Interspeech.2024-398},
  issn      = {2958-1796},
}

@misc{cho2024emosphere,
      title={EmoSphere++: Emotion-Controllable Zero-Shot Text-to-Speech via Emotion-Adaptive Spherical Vector}, 
      author={Deok-Hyeon Cho and Hyung-Seok Oh and Seung-Bin Kim and Seong-Whan Lee},
      year={2024},
      archivePrefix={arXiv},
      url={https://arxiv.org/abs/2411.02625}, 
}

@article{zhou2024emotional,
  title={Emotional Dimension Control in Language Model-Based Text-to-Speech: Spanning a Broad Spectrum of Human Emotions},
  author={Zhou, Kun and Zhang, You and Zhao, Shengkui and Wang, Hao and Pan, Zexu and Ng, Dianwen and Zhang, Chong and Ni, Chongjia and Ma, Yukun and Nguyen, Trung Hieu and others},
  journal={arXiv preprint arXiv:2409.16681},
  year={2024}
}

@article{jing2024enhancing,
  title={Enhancing Emotional Text-to-Speech Controllability with Natural Language Guidance through Contrastive Learning and Diffusion Models},
  author={Jing, Xin and Zhou, Kun and Triantafyllopoulos, Andreas and Schuller, Bj{\"o}rn W},
  journal={arXiv preprint arXiv:2409.06451},
  year={2024}
}

@article{tan2021survey,
  title={A survey on neural speech synthesis},
  author={Tan, Xu and Qin, Tao and Soong, Frank and Liu, Tie-Yan},
  journal={arXiv preprint arXiv:2106.15561},
  year={2021}
}

\end{document}